\documentclass[runningheads]{cl2emult}

\usepackage{makeidx}  
\usepackage{graphicx} 
\usepackage{subeqnar} 
\usepackage{multicol} 
\usepackage{cropmark} 
\usepackage{eso}      
\makeindex            

\def\kmsmpc{\,{\rm km\,s^{-1}\,Mpc^{-1}}}

\def\eblunits{\,{\rm nW\,m^{-2}\,sr^{-1}}}

\def\mden{\,{\rm M_\odot\,Mpc^{-3}}}
\def\lden{\,{\rm L_\odot\,Mpc^{-3}}}
\def\sfrd{\,{\rm M_\odot\,yr^{-1}\,Mpc^{-3}}}

\def\etal{{et al.\ }}

\def\spose#1{\hbox to 0pt{#1\hss}}
\def\lta{\mathrel{\spose{\lower 3pt\hbox{$\mathchar"218$}}
     \raise 2.0pt\hbox{$\mathchar"13C$}}}
\def\gta{\mathrel{\spose{\lower 3pt\hbox{$\mathchar"218$}}
	  \raise 2.0pt\hbox{$\mathchar"13E$}}}

\begin{document}
\title*{Cosmic Star Formation History and the Brightness of the Night Sky}
%
%

%
%
\titlerunning{Cosmic Star Formation History}
%
\author{Piero Madau}
\authorrunning{Piero Madau}
%
%
\institute{Institute of Astronomy, University of Cambridge, Cambridge
CB3 0HA, UK}

\maketitle              
\begin{abstract}
I review the constraints imposed by the observed extragalactic background 
light (EBL) on the history of the stellar birthrate in galaxies.
At faint magnitudes, the logarithmic slope of the galaxy counts 
is flatter than $0.4$ in all seven $UBVIJHK$ optical bandpasses of the
{\it Hubble Deep Field-South} imaging survey. The integration of the 
number counts provides a lower limit to the surface brightness
of the optical extragalactic sky of $15\,\eblunits$, comparable to
the intensity of the far-IR background from {\it COBE} data.
If the initial mass function has a Salpeter slope with a lower mass cutoff
consistent with observations of M subdwarf disk stars, a lower limit of
$\Omega_*>0.005\,I_{50}$ (at Hubble constant $50\,\kmsmpc$) is derived
for the visible (processed gas $+$ stars)
mass density needed to generate an extragalactic background
light (EBL) at a level of $50\,I_{50}\,\eblunits$. The current `best-guess' 
estimate to $\Omega_*$ is $0.012\,I_{50}$, about 16\% of the nucleosynthetic 
baryon density. The contribution of quasar
activity to the observed EBL is  unlikely to exceed 20\%.

\end{abstract}

\section{Introduction}

Recent progress in our understanding of faint galaxy data made possible
by the combination of {\it Hubble Space Telescope} {(\it HST}) deep imaging and
ground-based spectroscopy has dramatically increased our knowledge of the
evolution of the stellar birthrate in optically-selected galaxies from the
present-epoch up to $z\approx 4$ \cite{ref13}, \cite{ref33}, \cite{ref23}.
The explosion in the quantity of information available on the high-redshift
universe at optical wavelengths has been complemented by the measurement
of the far-IR/sub-mm background by DIRBE and FIRAS onboard the {\it COBE}
satellite \cite{ref21}, \cite{ref16}, \cite{ref30}, by the detection of 
distant ultraluminous sub-mm sources
with the SCUBA camera \cite{ref31}, \cite{ref11},
and by theoretical progress made in understanding how intergalactic gas follows
the dynamics dictated by dark matter halos until radiative, hydrodynamic, and
star formation processes take over \cite{ref2}, \cite{ref32}, \cite{ref28}.
The IR data have revealed the `optically-hidden' side of galaxy formation, and
shown that a significant fraction of the energy released by stellar
nucleosynthesis is re-emitted as thermal radiation by dust \cite{ref10}, 
\cite{ref20}. The underlying goal of all these efforts is to understand the 
growth of cosmic structures,
the internal properties of galaxies and their evolution, and ultimately
to map the star formation history of the universe from the end of
the cosmic `dark age' to the present epoch.

In this talk I will focus on the galaxy number-apparent magnitude
relation and its first moment, the integrated galaxy contribution to the
extragalactic background light (EBL). The
logarithmic slope of the differential galaxy counts is a remarkably simple
cosmological probe of the history of stellar birth in galaxies, as it must
drop below 0.4 to yield a finite value for the EBL. The recently released 
{\it Hubble Deep Field-South} (HDF-S) 
images, together with other existing {\it HST} and ground-based observations, 
provide a unique dataset to 
estimate the spectrum and amplitude of the optical EBL from discrete sources. 
Together with the
far-IR/sub-mm background, the optical EBL is an indicator of the total
luminosity of the universe, as the cumulative emission from young and evolved
galactic systems, as well as from active galactic nuclei (AGNs), is
recorded in this background. As such it provides, for a given initial mass
function,
a quantitative estimate of the baryonic mass that has been processed by
stars throughout cosmic history.

Unless otherwise stated, an Einstein-de Sitter (EdS) cosmology ($\Omega_M=1$, 
$\Omega_\Lambda=0$) with $H_0=100\,h\,\kmsmpc$ will be adopted in this talk.
All magnitudes will be given in the AB system. The work presented 
here has been done in collaboration with L. Pozzetti. 

\section{Galaxy counts}

The HDF-S dataset includes deep near-IR NICMOS images and the deepest 
observation ever made with the STIS 50CCD filterless imaging mode.
The galaxy sample used here was extracted from version 1 of the 
HDF-S catalog on ftp://archive. stsci.edu/pub/hdf\_south/version1/. 
At near-IR wavelengths (in the F110W, F160W, and F222M 
bandpasses, corresponding to the $J$, $H$, and $K$ filters), 
it consists of 425 objects detected in the $J+H$ image, over a field of
$50^{\prime\prime}\times 50^{\prime\prime}$.
The 50CCD (corresponding roughly to a $V+I$ filter) STIS catalog consists 
of 674 objects detected again over a field of the same size.

\begin{figure}
\centering
\vspace{-4cm}
\includegraphics[width=0.9\textwidth]{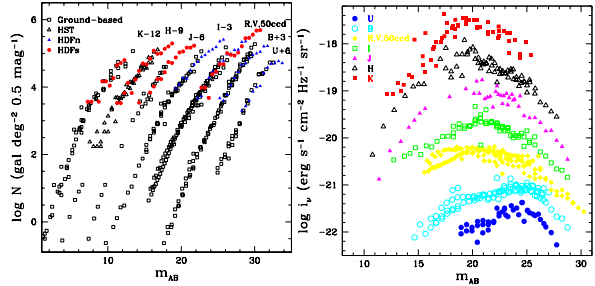}
\vspace{-5.5cm}
\caption[]{{\it Left:} Differential $UBVIJHK$ galaxy counts as a function of AB magnitudes.
The sources of the data points are given in the text. Note the decrease
of the logarithmic slope $d\log N/dm$ at faint magnitudes. {\it Right:} 
Extragalactic background 
light per magnitude bin,
$i_\nu=10^{-0.4(m_{\rm AB}+48.6)}N(m)$, as a function of $U$ ({\it
filled circles}), $B$ ({\it open circles}), $V$ ({\it filled pentagons}), $I$
({\it open squares}), $J$ ({\it filled triangles}), $H$ ({\it open triangles}), 
and $K$ ({\it filled squares}) magnitudes. For clarity, the $BVIJHK$ 
measurements have been multiplied by a factor of 2, 6, 15, 50, 150, and 600, 
respectively \cite{ref01}. 
}
\label{eps1}
\end{figure}

Figure 1 shows the HDF-N and -S galaxy counts compiled directly from the 
catalogs, as a function of AB 
isophotal magnitudes in the $UBVIJHK$ bandpasses for all galaxies 
with signal-to-noise ratio $S/N>3$
within the band. No correction for detection completeness have been made.
A compilation of 
existing {\it HST} and ground-based data is also shown \cite{ref01}, \cite
{ref29}. All magnitudes have been corrected to the AB system, while the 
second order
colour corrections for the differences in the filter effective wavelengths 
have not been applied to the ground-based data (for the typical
colours of galaxies in the HDF these corrections are less than 0.1 mag). 
The HDF optical counts agree well with previous surveys,
to within $20\%$ in the magnitude range $22<m_{\rm AB}<26$. 
One should note, however, that different algorithms used for `growing' the
photometry beyond the outer isophotes of galaxies can significantly change
the magnitude of faint galaxies. According to \cite{ref3}, roughly 
50\% of the flux from resolved galaxies with $V>23$ mag lie outside the 
standard-sized apertures used by photometric packages. An extragalactic
sky pedestal created by the overlapping wings of resolved galaxies may also 
contribute significantly to the sky level, and would be undetectable except 
by absolute surface photometry \cite{ref3}. Also, at faint magnitude levels, 
distant objects which are brighter than the nominal depth of 
the catalog may be missed due to the $(1+z)^4$ dimming factor.
All these systematic errors are inherent in {\it HST} faint-galaxy photometry;
as a result, our estimate of the integrated flux from resolved galaxies 
will typically be too low, and must be strictly considered 
as a {\it lower limit}. 

\section{The extragalactic background light}

The contribution of known galaxies to the optical EBL can be calculated 
directly by integrating the emitted flux times the differential number counts 
down to the detection threshold. I have used the compilation of ground-based, 
{\it HST}, and HDF data shown in Figure 1 to compute the integrated flux 
at $0.35\lta \lambda\lta 2.2\,\mu$m. 

In all seven bands, the slope of the differential number-magnitude relation is 
flatter than 0.4 above $m_{\rm AB} \sim 20$ (25) at near-IR (optical) 
wavelengths, and this flattening appears to be more pronounced at the shorter 
wavelengths.
The leveling off of the counts is clearly seen in Figure 1, where the function 
$i_\nu=10^{-0.4(m_{\rm AB}+48.6)}N(m)$ is plotted against apparent magnitude in
all bands. While counts having a logarithmic slope $d\log N/dm_{\rm AB}=
\alpha\ge0.40$ continue to add to the EBL at the faintest magnitudes, it
appears that the HDF survey has achieved the sensitivity to capture 
the bulk of the near-ultraviolet, optical, and near-IR extragalactic light 
from discrete sources. The flattening at faint 
apparent magnitudes cannot be due to the reddening of distant sources as 
their Lyman break gets redshifted into the blue passband,
since the fraction of Lyman-break
galaxies at (say) $B\approx 25$ is only of order 10\% \cite{ref34}.
Moreover, an absorption-induced loss of sources cannot explain the similar
change of slope in the $V,I,J,H,$ and $K$ bands.
While this suggests that the surface density of 
optically luminous galaxies is leveling off beyond $z\sim 1.5$, one should
worry about the possibility of a significant amount of light being missed 
at faint magnitudes.

\begin{figure}
\centering
\vspace{-2.5cm}
\includegraphics[width=.6\textwidth]{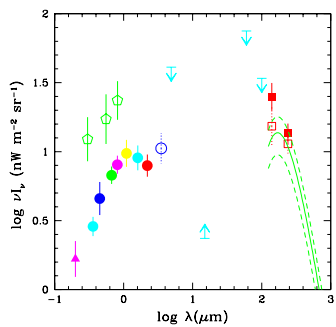}
\vspace{-2.5cm}
\caption[]{Spectrum of the optical extragalactic 
background light from resolved sources 
as derived from a compilation of ground-based and space-based galaxy counts 
in the $UBVIJHK$ bands ({\it filled dots}), together with the FIRAS 
125--5000 $\mu$m ({\it solid and dashed lines}) and DIRBE 140 and 
240 $\mu$m ({\it filled squares}) detections \cite{ref21}, \cite{ref16}. 
The {\it empty squares} show the DIRBE points after correction 
for WIM dust emission \cite{ref22}. Also plotted ({\it filled 
triangle}) is a FOCA-UV point at 2000 \AA\ \cite{ref1}, 
and a tentative detection at 3.5 $\mu$m ({\it empty dot}) from 
{\it COBE}/DIRBE observations \cite{ref9}. 
The empty pentagons at 3000, 5500, and 8000 \AA\ are recent detections from 
absolute photometry \cite{ref3}. Upper limits are from
\cite{ref21}, the lower limit from \cite{ref02}.
Note  that the 
values obtained by integrating the brightness of resolved galaxies are  
strict lower limits to the EBL intensity. 
}
\label{eps2}
\end{figure}

The spectrum of the optical EBL 
is shown in Figure 2, together with the recent results from {\it COBE}.
The value derived by integrating the galaxy counts down to very
faint magnitude levels (because of the flattening of the number-magnitude 
relation most of the contribution to the 
optical EBL comes from relatively bright galaxies) 
implies a lower limit to the EBL intensity in the 0.2--2.2 $\mu$m 
interval of $I_{\rm opt}\approx 15\,\eblunits$. Including the tentative 
detection at 3.5 $\mu$m by \cite{ref9} would boost 
$I_{\rm opt}$ to $\approx 19\,\eblunits$. 
Recent estimates of the optical EBL at 3000, 5500, and 8000 \AA\ 
from absolute surface photometry by \cite{ref3} lie between a 
factor of two to three higher than the integrated light from 
galaxy counts. Applying this correction factor to the range 3000--8000 \AA\
gives an optical EBL intensity in excess of $25\,\eblunits$ 
in the interval 0.2--3.5 $\mu$m. The {\it COBE}/FIRAS \cite{ref16}
measurements yield $I_{\rm FIR}\approx 14\,\eblunits$ in the 125--2000 $\mu$m 
range. When combined with the DIRBE \cite{ref21}, \cite{ref30}, \cite{ref22}
points at 140 and 240 $\mu$m, one gets a far-IR 
background intensity of $I_{\rm FIR}(140-2000\,\mu{\rm m})\approx 
20\,\eblunits$. 
The residual emission in the 3.5 to 140 $\mu$m region is poorly known, but it 
is likely to exceed $10\,\eblunits$ \cite{ref10}. A `best-guess' estimate to 
the total EBL intensity observed today is then
\begin{equation}
I_{\rm EBL}=55\pm 20\,\eblunits. 
\end{equation}
In the rest of my talk, I will adopt a reference value for the background light 
associated with star formation activity
over the entire history of the universe of $50\,I_{50}\eblunits$.  

\section{EBL from quasar activity}

A direct estimate of the contribution of quasars to the EBL 
depends on the poorly known bolometric correction and the possible 
existence of a distant population of dusty AGNs with 
strong intrinsic absorption, as invoked in many models for the X-ray 
background. These Type II QSOs, while undetected at 
optical wavelengths, could contribute significantly to the far-IR background.
It is in principle possible to bypass some of the above uncertainties by 
weighing the local mass density of black holes remnants \cite{ref03}.
Recent dynamical evidence indicates that supermassive black holes reside 
at the center of most nearby galaxies. The available data (see Fig. 3) 
show a correlation (but with a large scatter) between bulge and black 
hole mass, with $M_{\rm BH}\approx 0.006 \, 
M_{\rm sph}$ as a best-fit \cite{ref26}. The mass density in old spheroidal 
populations today is estimated to be
$\Omega_{\rm sph}h= 0.0018^{+0.0012}_{-0.00085}$ \cite{ref17}, implying a 
mean mass density of quasar remnants today    
\begin{equation}
\rho_{\rm BH}=3\pm 2\times 10^6\,h\, \mden.
\end{equation}
Since the observed energy density from all quasars is equal to the
emitted energy divided by the average quasar redshift, the 
total contribution to the EBL from accretion onto black holes is 
\begin{equation}
I_{\rm BH}={c^3\over 4\pi} {\eta \rho_{\rm BH}\over \langle 1+z\rangle}
\approx 4\pm 2.5\, \eblunits\, \eta_{0.05} {2.5\over \langle 1+z\rangle}
\end{equation}
($h=0.5$), where $\eta_{0.05}$ is the efficiency of accreted mass to radiation
conversion (in units of 5\%). Therefore, unless dust-obscured accretion onto 
supermassive black holes is a very efficient process ($\eta\gg 0.05$),  
a population of quasars peaking at $z\sim 1.5-2$ is expected to make a 
contribution to the brightness of the night sky not exceeding 10--20\%
\cite{ref15}, \cite{ref23}.

\begin{figure}
\centering
\includegraphics[width=.4\textwidth]{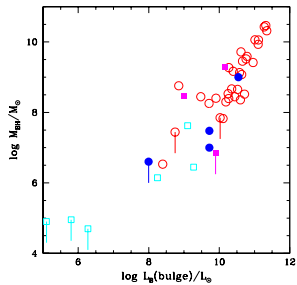}
\caption[]{Black hole mass distribution against the bulge luminosity of 
their host galaxies \cite{ref26}. Arrows indicate upper limits on $M_{\rm BH}$.
The symbols denote different galaxy types: {\it empty circles} (E), 
{\it filled squares} (S0), {\it filled circles} (Sab), and {\it empty 
squares} (Sbc-Scd).
}
\label{eps3}
\end{figure}

\section{The stellar mass density today}

With the help of some simple stellar population synthesis tools we can now 
set a lower limit to the total stellar mass density 
that produced the observed sky brightness and constrain 
the cosmic history of star birth in galaxies. One of the most serious 
uncertainties in this calculation has always been the lower cutoff 
(usually treated as a free parameter) of the 
initial mass function (IMF). 
Observations of M subdwarfs stars with the {\it HST} have recently
shed some light on this issue, showing that the IMF in the Galactic disk can 
be represented analytically
over the mass range $0.1<m<1.6$ (here $m$ is in solar units) by $\log 
\phi(m)={\rm const} -2.33 \log m -1.82(\log m)^2$ (\cite{ref19}, hereafter 
GBF). For $m>1$ this mass distribution agrees well with a Salpeter function. 
Observations of normal Galactic 
star-forming regions also show some convergence in the basic form of the 
IMF at intermediate and high
masses, a power-law slope that is consistent with the Salpeter value 
\cite{ref14}. 
In the following I will use a `universal' IMF with the 
GBF form for $m<1$, matched to a Salpeter slope for $m\ge 1$; the mass integral of this 
function is 0.6 times that obtained extrapolating a Salpeter function down to 
$0.1\, M_\odot$.\footnote{The bolometric light contributed by stars less 
massive than $1\,M_\odot$ is quite small for a `typical' IMF. The use of the 
GBF mass function at low masses instead of Salpeter leaves then the total 
radiated luminosity of the stellar population virtually unaffected.}

\begin{figure}
\centering
\vspace{-5cm}
\includegraphics[width=0.9\textwidth]{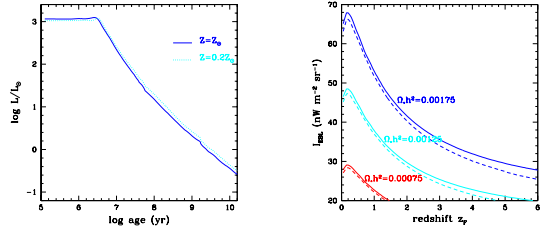}
\vspace{-5.5cm}
\caption[]{{\it Left:} Synthetic \cite{ref5}
bolometric luminosity versus age of a simple stellar 
population having total mass $M=1\,M_\odot$, metallicity $Z=Z_\odot$
({\it solid line}) and $Z=0.2\,Z_\odot$ ({\it dotted line}), and 
a GBF$+$Salpeter IMF (see text for details). 
{\it Right:} EBL observed at Earth from the instantaneous formation at 
redshift $z_F$ of a stellar population having the same IMF, solar metallicity,
and mass density $\Omega_*h^2=0.00175, 0.00125,$ and 0.00075, as a 
function of  
$z_F$. {\it Solid curves:} EdS universe with $h=0.5$. {\it Dashed curves:}  
$\Lambda$-dominated universe with $\Omega_M=0.3$, $\Omega_\Lambda=0.7$, and
$h=0.65$.
}
\label{eps4}
\end{figure}

As shown in Figure 4, the {\it bolometric} 
luminosity as a function of age $\tau$ of a simple stellar population (a single 
generation of coeval, 
chemically homogeneous stars having total mass $M$, solar metallicity, and the
above IMF) can be well approximated by 
\begin{equation}
L(\tau)= \left\{\begin{array}{ll} 1200\,L_\odot {M\over M_\odot} & 
\mbox{$\tau\le 2.6\,$ Myr;} \\
0.7\,L_\odot {M\over M_\odot} \left({\tau\over 1\,{\rm Gyr}}\right)^{-1.25} 
& \mbox{$2.6\le \tau\le 100\,$ Myr;} \\
2.0\,L_\odot {M\over M_\odot} \left({\tau\over 1\,{\rm Gyr}}\right)^{-0.8} & 
\mbox{$\tau>
100\,$ Myr} \end{array}
\right. \label{eq:bol}
\end{equation}
(cf \cite{ref7}).
Over a timescale of 13 Gyr (the age of the universe for an EdS cosmology with 
$h=0.5$), about 1.3 MeV per stellar baryon are radiated away. This number 
depends only weakly on the assumed metallicity of stars.    
In a stellar system with arbitrary star formation rate per comoving 
cosmological volume, $\dot \rho_*$, and formation epoch $t_F$, the 
comoving bolometric emissivity 
at time $t$ is given by the convolution integral
\begin{equation}
\rho_{\rm bol}(t)=\int_0^t L(\tau)\dot \rho_*(t-\tau)d\tau.
\label{eq:rhob}
\end{equation}
The total background light observed at Earth ($t=t_H$) is 
\begin{equation}
I_{\rm EBL}={c\over 4\pi} \int_{t_F}^{t_H} {\rho_{\rm bol}(t)\over 1+z}dt,
\label{eq:ebl}
\end{equation}
where the factor $(1+z)$ at the denominator is lost to cosmic expansion 
when converting from observed to radiated (comoving) luminosity density. 
To set a lower limit to the present-day mass density, $\Omega_*$, of 
processed gas $+$ stars 
(in units of the critical density $\rho_{\rm crit}=2.77 
\times 10^{11}\,h^2\mden$), consider now a scenario where all stars 
are formed {instantaneously} at redshift $z_F$.
The background light that would be observed at Earth from such an event 
is shown in Figure 4 as a function of $z_F$ for $\Omega_*h^2=0.00075, 
0.00125, 0.00175$ (corresponding to 9, 6.5, and 4 percent of the
nucleosynthetic baryon density, $\Omega_bh^2=0.0193\pm 0.0014$ \cite{ref6}),
and two different cosmologies. A couple of points are  
worth noting here: (1) the time evolution of the luminosity radiated by a 
simple stellar population makes
the dependence of the observed EBL from $z_F$ much shallower than the 
$(1+z_F)^{-1}$ 
lost to cosmic expansion, as the energy output from
stars is spread over their respective lifetimes; and (2) in order to generate 
an EBL at a level of $50\,I_{50}\,\eblunits$, one requires 
$\Omega_*h^2>0.00125\,I_{50}$ (for an EdS universe with $h=0.5$), hence a mean 
mass-to-blue light ratio today of $\langle M/L_B\rangle_*>3.5\,I_{50}$ 
for a present-day blue luminosity density of $\rho_B=2.0 \times 
10^8\,h\lden$ \cite{ref12}.
As shown in Figure 4, the dependence of these estimates on the 
cosmological model is rather weak.
With the adopted IMF, about 30\% of this mass will be returned to 
the interstellar medium in $10^8$ yr after intermediate-mass stars eject 
their envelopes and massive stars explode as supernovae. This `return fraction',
$R$, becomes 50\% after about 10 Gyr.

A visible mass density at the level of the above lower limit, $0.00125h^{-2}
\,I_{50}$,
while able to explain the measured sky brightness, requires all the stars that 
give origin to the observed light to have formed at $z_F\approx 0.2$, and is, 
as such, rather implausible. 
A more realistic
scenario appears to be one where the star formation density evolves as 
\begin{equation}
\dot\rho_*(z)=0.23\, {e^{3.4z}\over e^{3.8z}+44.7}\,\sfrd. \label{eq:mplot}
\end{equation}
This model fits reasonably well all 
measurements of the UV-continuum and H$\alpha$  luminosity densities 
from the present-epoch to $z=4$ after an extinction correction of
$A_{1500}=1.2$ mag ($A_{2800}=0.55$ mag) is applied to the data \cite{ref23},
and produce a total EBL of the right magnitude ($I_{50}=1$).  
Since about half of the present-day stars are formed at $z>1.3$
(hence their contribution to the EBL is redshifted away),  
the resulting visible mass density is $\Omega_*h^2=0.0031\,I_{50}$ ($\langle 
M/L_B\rangle_*=8.6\, I_{50}$). Note that this estimate ignores the recycling 
of returned gas into new stars.

The observed EBL therefore requires that between 7\% and 16\% 
of the nucleosynthetic baryons are today in the forms of stars, 
processed gas, and their remnants.
According to the most recent census of cosmic baryons, the mass density in 
stars and their remnants observed today is $\Omega_sh=0.00245^{+0.00125} _
{-0.00088}$ \cite{ref17}, corresponding to a mean visible mass-to-blue light 
ratio of $\langle M/L_B\rangle_s=3.4^{+1.7}_{-1.3}$ ($h=0.5$) (about 70\% of 
this mass is found in old spheroidal populations). While this is about a 
factor of 2.5 smaller than the visible mass density predicted by equation
(\ref{eq:mplot}), efficient recycling of ejected material into new star 
formation would tend to reduce the apparent discrepancy in the budget. 
Alternatively, the gas returned by stars may be ejected into the intergalactic
medium. With an IMF-averaged yield of returned metals of $y\approx 
1.5\,Z_\odot$,\footnote{Here we have taken $y\equiv \int mp_{\rm zm}\phi(m)dm 
\times [\int m\phi(m)dm]^{-1}$, the stellar yields $p_{\rm zm}$ of 
\cite{ref36}, and a GBF$+$Salpeter 
IMF.}\, the predicted mean metallicity at the present epoch is 
$y\Omega_*/\Omega_b=0.25\,Z_\odot$, in good agreement with the values
inferred from cluster abundances \cite{alvio}.

\clearpage
\addcontentsline{toc}{section}{Index}
\flushbottom
\printindex

\end{document}